\documentclass[twocolumn,showpacs,aps,prl,groupedaddress]{revtex4}

\usepackage{graphicx}
\usepackage{bm}

\begin{document}

\hsize\textwidth\columnwidth\hsize\csname@twocolumnfalse\endcsname

\title{Qubit decoherence by low-frequency noise}

\author{K. Rabenstein, V.A. Sverdlov, and D.V. Averin}
\address{Department of Physics and Astronomy, Stony Brook University,
SUNY, Stony Brook, NY 11794-3800}

\date{\today}

\begin{abstract}
We have derived explicit non-perturbative expression for
decoherence of quantum oscillations in a qubit by low-frequency
noise. Decoherence strength is controlled by the noise spectral
density at zero frequency while the noise correlation time $\tau$
determines the time $t$ of crossover from the $1/\sqrt{t}$ to the
exponential suppression of coherence. We also performed Monte
Carlo simulations of qubit dynamics with noise which agree with
the analytical results and show that most of the conclusions
are valid for both Gaussian and non-Gaussian noise.

\end{abstract}

 \maketitle

Despite the large number of successful demonstrations of coherent
quantum oscillations in individual \cite{b1,b2,b3,b4,b5,b6,b7,b8}
and coupled \cite{b9} Josephson-junction qubits, quantitative
understanding of these oscillations is so far limited. The main
area of discrepancy between experiment and theory is qubit
decoherence. The typical quality factors of reported oscillations,
while not as large as required by potential applications in
quantum computation, are still quite large in physics term
(typically not less than $20\div 30$). This fact should imply
weak decoherence describable by the standard perturbation theory
in qubit-environment coupling (see, e.g., \cite{b10}).
Several basic features of this theory, however, do not agree with
experimental observations. Most importantly, observed decay time
$T_2$ of coherent oscillations is typically shorter than the
energy relaxation time $T_1$ even at optimal qubit bias points
\cite{b4,b3,b12} where perturbation theory predicts no pure
dephasing terms. Another discrepancy is between the observed
two-qubit decoherence rate \cite{b9} and its values that can be
obtained from the perturbation theory under natural assumptions
\cite{b13}.

Qualitatively, the basic reason for discrepancy between $T_1$
and $T_2$ is the low-frequency noise that can reduce $T_2$
without changing significantly the relaxation rates. Mechanisms
of low-frequency, or specifically $1/f$, noise exist in all
solid-state qubits: background charge
fluctuations for charge-based qubits \cite{b14}, impurity
spins or trapped fluxes for magnetic qubits \cite{b15}.
Manifestations of this noise are observed in the echo-type
experiments \cite{b12}. Low-frequency noise for qubits is
also created by the electromagnetic fluctuations in filtered
control lines.

The goal of our work is to develop quantitative theory of
low-frequency decoherence by studying qubit dynamics under the
influence of noise with small characteristic amplitude $v_0$
and long correlation time $\tau$. In the case of Gaussian noise,
we obtained explicit non-perturbative expression describing
decay in time of coherent qubit oscillations. The strength of
decoherence in this expression is controlled by the noise
spectral density at zero frequency, $v_0^2\tau$. For long
correlation times $\tau \gg \Delta^{-1}$, where $\Delta$ is
the qubit tunnel amplitude, the spectral density $v_0^2\tau$
can be large even for weak noise $v_0 \ll \Delta$ and our
analytical results are exact as function of $v_0^2\tau$ in this
limit. We also performed direct numerical simulations of the
low-frequency qubit decoherence. The simulation results
confirm analytical expressions for Gaussian noise and show that
our main conclusions:  cross-over from the $1/\sqrt{t}$ to the
exponential suppression of coherence at time $t\simeq \tau$;
and the strength of decoherence controlled by the noise
spectral density $S_v(0)$ at zero frequency remain valid for
non-Gaussian noise. The decoherence rates for Gaussian and
non-Gaussian noise behave, however, quite differently as
functions of $S_v(0)$. In physics terms, the two noise models we
study correspond, respectively, to low-frequency electromagnetic
noise, and noise of localized excitations in the situation when
only a few excitations with comparable time scales are coupled
to the qubit.

The Hamiltonian of a qubit with a fluctuating bias energy $v(t)$
(see inset in Fig.\ 1b) is:
\begin{equation}
H= -\frac{1}{2}[\Delta\sigma_{x}+(\varepsilon+v(t)) \sigma_{z}]
\, ,
\label{e1} \end{equation}
where $\varepsilon$ is the average bias, and $\sigma$'s here
and below denote Pauli matrices. In this work, we mostly focus
on the situation when the noise $v(t)$ has characteristic
correlation time $\tau$, i.e., the noise correlation function
and its spectral density can be taken as
\begin{equation}
\langle v(t) v(t') \rangle = v_0^2 e^{-|t-t'|/\tau} \, ,
\;\;\; S_v(\omega) = \frac{2 v_0^2 \tau}{1+(\omega \tau)^2}
\, ,
\label{e2} \end{equation}
where $v_0$ is the typical noise amplitude and $\langle ...
\rangle$ denotes average over different realizations of noise.
We assume that the temperature $T$ of the noise-producing
environment is large on the scale of the cut-off frequency
$1/\tau$, and it can be treated as classical. (In the regime
of interest, $1/\tau \ll \Delta$, the temperature can
obviously be still small on the qubit energy scale.)

The two effects of the weak noise on the dynamics of the qubit
(\ref{e1}) are the transitions between two energy eigenstates
with energies $\pm \Omega /2$, $\Omega \equiv (\Delta^2 +
\varepsilon^2)^{1/2}$, and ``pure'' (unrelated to transitions)
dephasing that suppresses coherence between these states.
Within the standard perturbation theory, the transition rate
is proportional to $S_v(\Omega)=v_0^2/\Omega^2 \tau$. One can
see that the
condition of weak noise $v_0\ll \Delta$ makes the transition
rate small compared both to $\Delta$  and $1/\tau$ ensuring
that the perturbation theory is sufficient for the description
of transitions. As discussed qualitatively in the introduction,
the fact that the noise correlation time is long, $\tau \gg
\Delta^{-1}$, makes the perturbation theory inadequate for the
description of pure dephasing. For low-frequency noise, a
proper (non-perturbative in $v_0^2\tau$) description is
obtained by looking at the accumulation of the noise-induced
phase between the two instantaneous energy eigenstates.
If $v_0\ll \Delta$, one can determine the rate of
accumulation of this phase by expanding the energies in
noise amplitude $v(t)$. Also, in this case the dephasing
rate is larger than the transition rate and can be calculated
disregarding the transitions. The factor $F(t)$ describing
suppression in time of coherence
between the two states (i.e., suppression of the off-diagonal
element $\rho_{12}$ of the qubit density matrix in the energy
basis: $\rho_{12} (t) = F(t)\rho_{12} (0)e^{-i\Omega t}$)
can be written then as follows:
\begin{equation}
F(t)= \langle \exp \{-i \int_0^t [\frac{\varepsilon v(t')}{
\Omega} + \frac{\Delta^2 v^2(t')}{2\Omega^3} ]dt'\} \rangle
\, .
\label{e3} \end{equation}

For {\em Gaussian noise}, the correlation function (\ref{e2})
determines the noise statistics completely. In this case, it
is convenient to take the average in Eq.\ (\ref{e3}) by writing
it as a functional integral over noise. For this purpose, and
also for use in the numerical simulations, we start with the
``transition'' probability
$p(v_1,v_2,\delta t)$ for the noise to have the value $v_2$ a
time $\delta t$ after it had the value $v_1$:
\begin{eqnarray}
p(v_1,v_2,t)= [2\pi v_0^2(1-e^{-2\delta t/\tau})]^{-1/2} \times
\nonumber \\
\exp \{-\frac{1}{2v_0^2}\frac{(v_2-v_1e^{-\delta t/\tau})^2
}{1-e^{-2\delta t/\tau}} \} \, .
\label{ea} \end{eqnarray}
Using this expression we introduce the probability of
specific noise realization as $p_0(v_1)\cdot p(v_1,v_2,\delta
t_1)\cdot p(v_2,v_3,\delta t_2)\cdot ... $, where
$p_0(v)=(2\pi v_0^2)^{-1/2} \exp \{-v^2/2v_0^2 \}$ is the
stationary Gaussian probability distribution of $v$. Taking the
limit $\delta t_j \rightarrow 0$ we see that the average over
the noise can be written as the following function integral:
\begin{eqnarray}
\langle ...  \rangle= \int dv(0) dv(t) Dv(t') ... \nonumber \\
\times \exp \left\{ -\frac{v(0)^2+v(t)^2}{4v_0^2} -\frac{1}
{4v_0^2 \tau} \int_0^t dt' (\tau^2 \dot{v}^2+v^2) \right\} \, .
\label{e4} \end{eqnarray}
Since the average in Eq.\ (\ref{e3}) with the weight (\ref{e4})
is now given by the Gaussian integral, it can be calculated
straightforwardly:
\begin{eqnarray}
F(t)= F_0(t) \exp\left[-\alpha^2\left(\frac{\nu t}{\tau}-
2[\coth \frac{\nu t}{2\tau}+\nu]^{-1} \right)\right],
\label{e5} \\
F_0(t)= e^{t/2\tau} [\cosh(\nu t/\tau)+ \frac{1+\nu^2}{2\nu}
\sinh(\nu t/\tau)]^{-1/2} \, , \nonumber
\end{eqnarray}
where $\nu \equiv \sqrt{1+2 i v_0^2\Delta^2\tau/\Omega^3}$
and $\alpha \equiv \varepsilon\tau v_0/\Omega\nu^{3/2}$.

\begin{figure}[htb]
\includegraphics[scale=0.4]{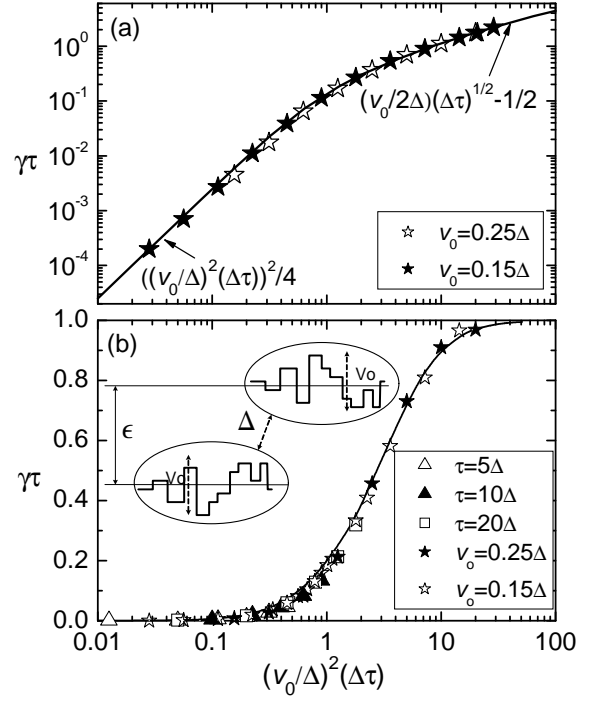}
\caption{The rate $\gamma$ of exponential qubit decoherence
at long times $t\gg\tau$ for $\varepsilon =0$ and (a) Gaussian
and (b) a model of the non-Gaussian noise with characteristic
amplitude $v_0$ and correlation time $\tau$. Solid lines give
analytical results: \protect Eq.\ (\ref{e7}) in (a) and Eq.\
(\ref{e16}) in (b). Symbols show $\gamma$ extracted from
Monte Carlo simulations of qubit dynamics. Note different
scales for $\gamma$ in parts (a) and (b). Inset in (b) shows
schematic diagram of qubit basis states fluctuating under the
influence of noise $v(t)$.}
\label{fig1} \end{figure}

Equation (\ref{e5}) is our main analytical result for dephasing
by the Gaussian noise. To analyze its implications, we start
with the case $\varepsilon =0$, where pure qubit dephasing
vanishes in the standard perturbation theory. Dephasing
(\ref{e5}) is still non-vanishing and its strength depends on
the noise spectral density at zero frequency $S_v(0)=2 v_0^2
\tau$ through $\nu= \sqrt{1+is}$, $s\equiv S_v(0)/\Delta$.
For small and large times $t$ Eq.\ (\ref{e5}) simplifies to:
\begin{equation}
F(t)= \left\{\begin{array}{cc} \displaystyle
{\left[\frac{1+t/\tau}{1+t/\tau+ist/2\tau}\right]^{1/2}},
& t \ll \tau\, , \\
2\sqrt{\nu}e^{-(\gamma+i\delta)t}/(1+\nu)\, , & t \gg \tau \, ,
\end{array} \right.
\label{e6} \end{equation}
where
\begin{equation}
\gamma= \frac{1}{2\tau} \left[ \left( \frac{(1+s^2)^{1/2}+1}{2}
\right)^{1/2} -1 \right]\, .
\label{e7} \end{equation}
Besides suppressing coherence, the noise also shifts the
frequency of qubit oscillations. The corresponding frequency
renormalization is well defined for $t\gg \tau$:
\begin{equation}
\delta = \frac{1}{2\tau} \left[\frac{(1+s^2)^{1/2}-1}{2}
\right]^{1/2} \, .
\label{e8} \end{equation}

Suppression of coherence (\ref{e6}) for $t\ll \tau$ can be
qualitatively understood as the result of averaging over the static
distribution of noise $v$. In contrast to this, at large times
$t\gg \tau$, the noise appears to be $\delta$-correlated, the fact
that naturally leads to the exponential decay (\ref{e6}). This
interpretation means that the two regimes of decay should be
generic to different models of the low-frequency noise. Indeed,
they exist for the non-Gaussian noise considered below, and
are also found for Gaussian noise with $1/f$ spectrum \cite{b16}.
Crossover between the two regimes takes place at $t\simeq \tau$,
and the absolute value of $F(t)$ in the crossover region can
be estimated as $(1+s^2)^{-1/4}$, i.e. $s$ determines
the amount of coherence left to decay exponentially. The rate
(\ref{e7}) of exponential decay shows a transition from the
quadtratic to square-root behavior as a function of $S_v(0)$
that can be seen in Fig.\ 1a, which also shows the decay rate
extracted from numerical simulations of Gaussian noise.
(Numerical procedure is discussed below.) One can see that our
analytical and numerical results agree well for quite large
noise amplitude $v_0$.

Non-zero qubit bias $\varepsilon$ leads to additional dephasing
$F(t)/F_0(t)$ described by the last exponential factor in Eq.\
(\ref{e5}). One can see that similar to $F_0(t)$ additional
dephasing exhibits the crossover at $t\simeq \tau$ from
``inhomogenious broadening'' (averaging over the static
distribution of the noise $v$) to exponential decay at $t\gg
\tau$. In contrast to $F_0(t)$, the short-time decay is now
Gaussian:
\[ \ln \big[\frac{F(t)}{F_0(t)} \big] = -
\frac{\varepsilon^2}{\Omega^2} \cdot \left\{\begin{array}{cc}
\displaystyle v_0^2 t^2/2 \, , &  t \ll \tau\, , \\
v_0^2 \tau t/(1+is(\Delta/\Omega)^3) \, , & t \gg \tau \, .
\end{array} \right.  \]
We see that, again, the rate of exponential decay depends
non-trivially on the noise spectral density $S_v(0)$, changing
from direct to inverse proportionality to $S_v(0)$ at small
and large $s$, respectively.

Our approach can be used to calculate the rate of exponential
decay at large times $t$ for Gaussian noise with arbitrary
spectral density $S_v(\omega)$. Such a noise can be
represented as a sum of noises (\ref{e2}) and appropriate
transformation of variables in this sum enables one to write
the average over the noise as a functional integral similar
to (\ref{e4}). For calculation of the
relaxation rate at large $t$, the boundary terms in the
integral (\ref{e4}) can be neglected and it is dominated by
the contribution from the ``bulk'' which can be conveniently
written in terms of the Fourier components
\[ v_n=(2/t)^{1/2} \int_0^t dt' v(t')\sin \omega_nt'\, , \;\;\;
\omega_n=\pi n/t \, .\]
Then,  $\langle ...  \rangle= \int Dv ... \exp \{ -(1/2)
\sum_n |v_n|^2/S_v(\omega_n) \} $.
Combining this equation and Eq.\ (\ref{e3}) we get at large $t$:
\begin{eqnarray}
F(t)=  \exp \left\{- \frac{t}{2} \left[ \frac{\varepsilon^2\Omega
S_v(0) }{\Omega^3+iS_v(0)\Delta^2}+  \right. \right.\nonumber
\\ \frac{1}{\pi}
\left. \left. \int_0^{\infty} d \omega \ln (1+iS_v(\omega)
\Delta^2/\Omega^3 ) \right] \right\} \, .
\label{e12} \end{eqnarray}
For unbiased qubit, $\varepsilon=0$, this equation coincides with
the one obtained by more involved diagrammatic perturbation theory
in quadratic coupling \cite{b16}.

\begin{figure}[htb]
\includegraphics[scale=0.4]{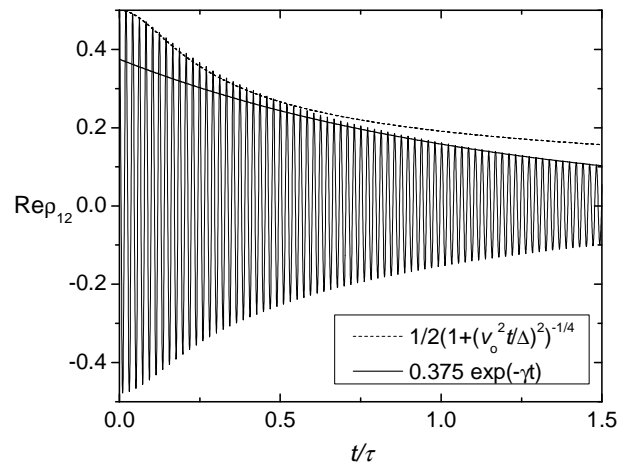}
\caption{The profile of coherent quantum oscillations in
an unbiased qubit dephased by the non-Gaussian noise with
characteristic amplitude $v_0=0.15\Delta $ and correlation
time $\tau=300\Delta^{-1}$ obtained by direct simulation of
qubit dynamics with noise. Solid line is the exponential
fit of the oscillation amplitude at large times. Dashed line
is the initial $1/\sqrt{t}$ decay caused by effectively
static distribution of $v$.}
\label{fig2} \end{figure}

To check how well the analytic theory described above works for
finite noise amplitude $v_0$, and to see how sensitive the results
are to the assumption of the Gaussian noise, we performed
Monte Carlo simulations of the qubit oscillations under the
influence of Gaussian and non-Gaussian noise. We looked
specifically at the coherent oscillations of a qubit with
Hamiltonian (\ref{e1}) that start in one of the eigenstates of the
$\sigma_{z}$ operator, focusing on the case $\varepsilon=0$.
The qubit density matrix was averaged over up to $10^7$
realizations of noise. In the case of Gaussian noise,
realizations were built using the transition probability
(\ref{ea}). For non-Gaussian noise we used the model of Ref.\
\onlinecite{b17}, which should provide an appropriate description
of the situation when a qubit is coupled to several fluctuators
with similar characteristic time scale $\tau$ of the fluctuations.
In this model the fluctuators create random qubit bias $v$ which
remains constant for some (random) time interval after which it
is updated and the new value remains constant during the next time
interval, etc. The time intervals between bias updates are taken
to be distributed according to the Poisson distribution with
characteristic time $\tau$. For more direct comparison with the
Gaussian noise, we assumed Gaussian distribution $p_0(v)$ of $v$.
The correlation function of $v(t)$ defined in this way is given
by the same Eq.\ (\ref{e2}).

Example of the oscillations dephased by such a noise is given
in Fig.\ 2. It shows real part of the
off-diagonal element $\rho_{12} (t)$ of the qubit density matrix
in the energy eigenstates basis. For oscillations starting in
one of the $\sigma_z$ eigenstates, $\rho_{12} (0)=1/2$.
Similarly to the case of Gaussian noise, we consider only weak
noise, $v_0\ll \Delta$. In this case, there is a crossover at
$t\simeq \tau$ in the oscillation amplitude from the initial
$1/\sqrt{t}$ suppression of coherence due to averaging over
static potential distribution:
$\rho_{12} (t) =\rho_{12} (0)/(1+iv_0^2t/\Delta)^{1/2} $
(neglecting all terms of order $t/\tau$), to exponential
suppression at $t\gg \tau$.

The rate of the exponential decay can be found analytically
as follows. Expansion of the average qubit density matrix
$\rho(t)$ in the number of ``jumps'' of $v(t)$ leads to
the Dyson-like equation for its evolution \cite{b17}:
\begin{eqnarray}
\rho(t) =e^{-t/\tau} \langle S(t,0)\rho(0)S^{\dagger}(t,0)
\rangle     \nonumber \\
+\int_{0}^{t}\frac{dt'}{\tau}e^{-(t-t')/\tau} \langle S(t,t')
\rho(t')S^{\dagger}(t,t') \rangle \, ,
\label{e14}\end{eqnarray}
where $\langle ... \rangle$ denotes the average over the
distribution of $v$. For weak noise, introducing
slowly-varying amplitude $r$ of $\rho_{12} (t)=r(t)
e^{-i\Delta t}$, one can reduce Eq.\ (\ref{e14}) to the
equation for $r(t)$ neglecting rapidly oscillating terms:
\begin{eqnarray}
r(t) =e^{-t/\tau}r(0)(1+iv_0^2t/\Delta)^{-1/2} \nonumber \\
+\int_{0}^{t}\frac{dt'}{\tau} e^{-(t-t')/\tau}r(t')
(1+iv_0^2(t-t')/\Delta)^{-1/2}  \, .
\label{e15} \end{eqnarray}
With the exponential ansatz for $r(t)$: $r(t)\propto
e^{-(1-\lambda)t/\tau}$ Eq.\ (\ref{e15}) gives then
equation for the parameter $\lambda$:
\begin{equation}
\lambda =\int_0^\infty dx e^{-x} \left[1+\frac{ix s}{2\lambda}
\right]^{-1/2}\, .
\label{e16} \end{equation}
(Omission of the first term in Eq.\ (\ref{e15}) is justified
by the final result for the oscillation decay rate
$\gamma =(1-\mbox{Re} [\lambda])/\tau$.) Asymptotics of
$\gamma$ found from Eq.\ (\ref{e16}) are:
\begin{equation}
\gamma = \frac{1}{\tau} \times
\left\{ \begin{array}{cc} s^2/8 , & s \ll 1\, ,  \\
1-16\pi/s^2, & s \gg 1 \, .
\label{e17} \end{array} \right.  \end{equation}

The rate $\gamma(s)$ evaluated from Eq.\ (\ref{e16}) is shown
in Fig.\ 1b, together with the pure dephasing rates found
numerically by fitting the oscillation amplitude
(similar to that shown in Fig.\ 2) at $t\gg \tau$ and
subtracting the contribution $S_v(\Delta)$ to dephasing from
real transitions. One can see that Eq.\ (\ref{e16}) indeed
gives an accurate description of pure dephasing rates.
Similarly to the case of Gaussian noise, $\gamma$
depends only on $S_v(0)$. In both situations, $\gamma \propto
v_0^4\tau/\Delta^2$ for small $v_0^2\tau/\Delta$, the fact
that can be explained by the lowest-order perturbation theory
in qubit energy fluctuations. In the non-perturbative regime,
however, the behavior of $\gamma$ as function of $S_v(0)$ is
model-dependent and varies from saturation (\ref{e17}) to
$\sqrt{S_v(0)}$-growth (\ref{e7}).

In summary, we developed non-perturbative theory of qubit
dephasing within two models of Gaussian and non-Gaussian
low-frequency noise and performed Monte Carlo simulations
of qubit dynamics within these models. The theory agrees well
with simulations and shows that the decoherence strength is
controlled by the noise spectral density at zero frequency.
It allows for generalizations in several
experimentally-relevant directions and should be useful for
analysis of observed shapes of quantum qubit oscillations.

\vspace*{.4ex}

This work was supported in part by ARDA and DOD under the DURINT
grant \# F49620-01-1-0439 and by the NSF under grant \# 0325551.
The authors would like to thank T. Duty, K. Likharev, J. Lukens,
Yu. Makhlin, Y. Nakamura, Yu. Pashkin, and A. Schnirman for
useful discussions.

\end{document}